# Network Effects Research: A Systematic Review of Theoretical Mechanisms and Measures


Alejandra Medina[1], Michael D. Siciliano[1], Weijie Wang[2], and Qian Hu[3]

[1] University of Illinois at Chicago, College of Urban Planning and Public Affairs, Department of Public Administration
[2] University of Missouri, Harry S Truman School of Public Affairs
[3] University of Central Florida, School of Public Administration



This article contributes to the network effectiveness literature by identifying the theoretical mechanisms and network measures scholars in public administration and policy use to draw inferences between network structures and network effects. We conducted a systematic review of empirical network effects research in 40 journals of public administration and policy from 1998 to 2019. We reviewed and coded 89 articles and described the main social theories used in the network effectiveness literature and the associated mechanisms that translate network structures to network effects. We also explain how scholars operationalize those theoretical mechanisms through network measures. Overall, our findings reflect that there is a limited use of social theories for the explanation of network effects and in some cases, there is an inconsistent use of network measures associated with theories. Moreover, we identify the main challenges related to network effects, including the difficulty of isolating specific mechanisms related to a particular social theory, the use of network structures both as a mechanism and as a measure, and the lack of data to examine network dynamics and coevolution.




**Introduction**

Networks have long been viewed as a suitable form of organizing to address complex policy problems and coordinate public service delivery (Provan & Milward, 2001). However, much of the empirical research on the effectiveness of networks in public settings did not begin until the 1990s. Since that time, network research has grown tremendously. There exist relevant frameworks to assist researchers in assessing network performance (Provan, K. & Milward, 1995; Raab et al., 2015; Herranz, 2010) along with syntheses of the extant network effectiveness literature (Cristofoli & Markovic, 2016; Kenis & Provan, 2009; Turrini et al., 2010). Despite this growth, there remains a lack of understanding as to how network structures determine outcomes. Therefore, it is important to identify the specific mechanisms associated with different relational configurations that produce outcomes for the network members or the whole network (Hedström, 2008).

Provan and Milward (2001) identified two critical challenges confronting research on network effectiveness: i) how to reach consensus on network goals and performance metrics, and ii) how to identify the primary aspects of networks that may influence performance. Our study is concerned with the latter challenge. We conduct a systematic review of the network literature in public administration and policy to identify the theoretical mechanisms and network measures scholars use to draw inferences between network structures and network effects.[1] Our systematic review covers network research for 21 years (1998-2019) in 40 journals of public administration and policy.

This article contributes to the network effectiveness literature in three ways. First, we provide a concise description of the theories and the associated mechanisms that translate network structures into network effects. This description and review distill the primary



mechanisms by which our field connects structure to performance. Overall, we find that there is limited use of social theories for the explanation of network effects. Second, we examine how scholars operationalize those mechanisms through different nodal and network measures. We find an inconsistent use of measures associated with specific social theories as scholars often use the same network measure to operationalize different theoretical mechanisms. We further analyze which network measures are most closely associated with particular mechanisms. Third, we identify the main challenges related to the extant research on network effectiveness. These challenges include the difficulty of isolating particular mechanisms when multiple mechanisms are associated with a given theory or outcome and the lack of suitable data to study network dynamics and the coevolution of network structure and effects. In the following section we explain the methodology used for the systematic review and then discuss the main theories and network measures used in the network effects literature.

## Systematic review: Methodology

To examine the application of network theories of performance in public administration and policy, we reviewed empirical articles about public sector networks published from January 1998 to May 2019. To identify relevant articles, we followed the slightly modified PRISMA protocol (Moher et al., 2009) used by Siciliano et al. (2021) and Kapucu et al. (2017). We conducted a general search in 40 main journals of public administration and policy.[2] Our search and exclusion process consisted of five steps. First, we searched on the websites of each journal for articles that included in the title, abstract, or keywords the following terms "network," "network analysis," "collaboration," and "collaborative." In this first step, we found 2,402 articles that met our criteria. Then, we reviewed the abstracts of these articles to ensure that they were about network studies and not using the word network as a metaphor and we kept 1,062 articles. Third,



we reviewed the articles with particular emphasis on the methodology section to verify that the authors used social network analysis. A total of 282 articles met this criterion. Fourth, we confirm the articles were about public sector networks and removed studies concerning private networks (e.g., articles about engineering firms, the airline industry, or high-tech companies); 196 articles met this criterion. Finally, since we were interested in understanding the effect of networks, we removed 107 articles that use the network as the dependent variable and analyze network formation processes and thus kept only articles about network effectiveness. We found 89 articles that met this criterion.

For the analysis of the 89 articles that met our search and exclusion criteria, we designed a comprehensive coding protocol to extract from the articles specific information on the type and number of nodes in the network, type of ties, method of network data collection, primary unit of analysis, area of study, research method, each network related hypothesis and its associated use of theory and measures, and the different factors identified by the authors as drivers for performance.

After coding, we removed an additional 15 articles from further analysis due to a lack of network effect data. Even though these 15 articles considered the network as an independent variable to predict network effects, the text of the article contained insufficient information regarding the variables of interest included in our coding protocol. In this regard, the final number of articles included in the analysis was 74.

Since we were interested in identifying the use of theories and network measures in the network effects literature, we open coded the exact text of each hypothesis, the theoretical mechanism used by the authors to frame and support the hypothesis, the authors' own description of the mechanism at work in the hypothesis, and the network measures used to



operationalize the hypothesis. Based on the findings of the open coding process, we identified a set of 10 theoretical mechanisms, 5 structural arguments, and 21 distinct network measures used by the authors. Once we identified these categories, each hypothesis was recoded into its appropriate categories.

Next, we designed a pilot test where the four authors coded 10 articles to compare and develop consistency in coding and adjust the coding protocol where needed. Once we concluded the pilot test and finalized the coding protocol, we created two teams with two coders each and coded all articles. For the calculation of interrater reliability, we calculated the percentage agreement measure and got a rate of 82.35%. All disagreements between coders were discussed by the entire team, and consensus was reached on all coding decisions.

The specificity of the coding protocol allows us to analyze the results at the article and hypothesis level. Eighteen out of the 74 articles were descriptive papers. While these articles did not conduct inferential tests, we kept the articles in the analysis as they included a clear description of the relationship between network attributes and network effects. For instance, one descriptive article analyzed the impact of heterogeneity on network effectiveness (Varda & Retrum, 2015). The authors operationalized heterogeneity by analyzing the diversity of the network members. Figure 1 reported the number of eligible articles and the number of relevant hypotheses extracted from the inferential studies.

Figure 1. Inferential and Descriptive Articles Included in the Analysis

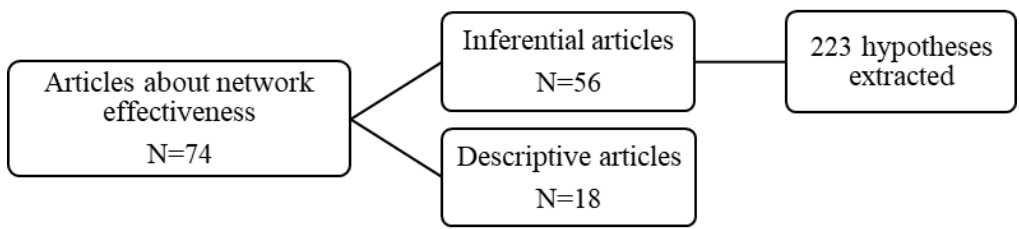

## Findings: Theories and Measures in the Network Effectiveness Literature

We identified and coded 223 hypotheses across the 56 inferential articles. Each hypothesis was coded based on the theory used by the authors as well as the network measures used to operationalize the theory. However, there were cases where the authors did not rely on a particular theory or mechanism to justify their hypothesis regarding network structure and performance but rather relied on the general network literature and thus offered more structural arguments. For example, some scholars hypothesized that a higher number of collaboration partners is associated with better performance and operationalize the number of collaboration ties through degree centrality scores. Building on Contractor, Wasserman, and Faust (2006), we grouped the different theories or justifications we identified into two main categories: social theories and structural arguments. According to Coleman (1994), the main focus of social theories is the explanation of social systems, and "this could be as small as a dyad or as large as a society or a world system" (p.2). The category of *social theories* includes contagion, resource dependence, collective action, exchange, and homophily theories, while the *structural arguments* focus on the effect of network centralization, density, or embeddedness. Another distinction between these two categories is that the social theories are based on mechanisms that are generally exogenous to the network or can operate in non-network settings, while the structural arguments emphasize mechanisms that are endogenous to the network (Contractor et al., 2006; Siciliano et al., 2021). Table 1 shows the different theories and arguments we identified in the eligible articles.



Table 1. Categorization of Theories

| | Theory |
|---|---|
| Social Theories | Collective Action |
| | Heterophily |
| | Homophily |
| | Policy Diffusion/Social Influence |
| | Resource Dependence |
| | Social Capital-Bridging |
| | Social Capital-Bonding |
| | Social Capital-Generic |
| | Social Capital - Trust |
| | Social Exchange |
| Structural Arguments | Structural Argument-Centrality |
| | Structural Argument-Density |
| | Structural Argument-Structural Embeddedness |
| | Structural Argument-Structural Equivalence |
| | Structural Argument-Transitivity |
| Other | Multiple |
| | Other |
| | NA |

Figure 2 shows the number of times these different theories and structural arguments were used in the network effectiveness literature. We found that the majority of the hypotheses relied on a select few theories to predict network effects. The most prominent ones are social capital-bridging and structural arguments around centrality. Approximately 2% of the hypotheses were coded as "Multiple" when the authors relied on more than one theory for the justification. Nearly 4% were coded as "Other" when scholars relied on theories that were not commonly used and only were found in one hypothesis like production theory, systems theory, and entropy theory. In addition, 40% of the hypotheses were coded as "NA" in the theory category because the authors did not rely on either social theories or structural arguments for the hypothesis framing. We will discuss this category further in the findings section.





*Figure 2. Use of Social and Structural Theories Across Hypotheses*

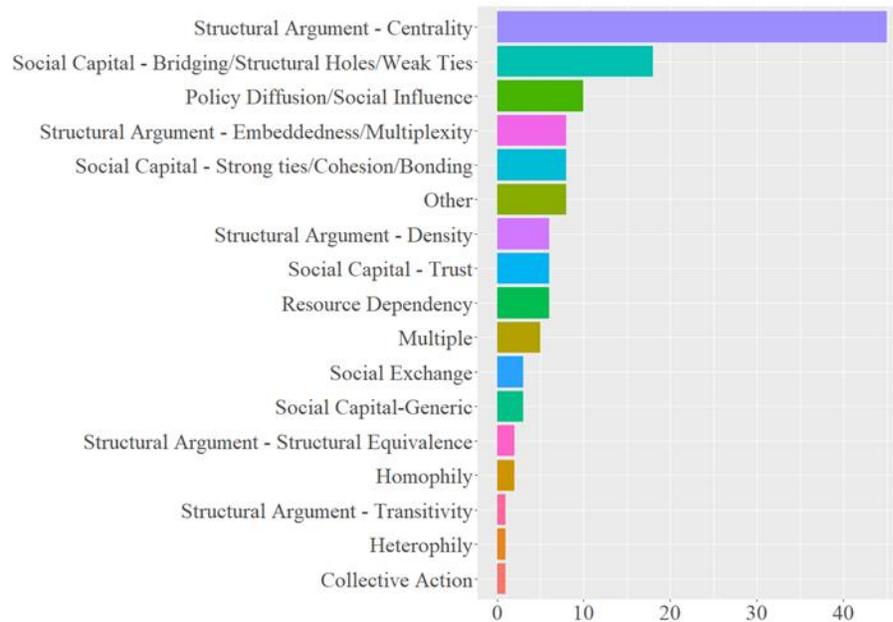

In addition to the identification of theory use, we also coded the specific network measures used by the authors to capture a particular mechanism along with the rationale supporting the use of that measure. We find that scholars use network measures for the operationalization of different variables posited to impact network effectiveness not only in inferential studies but also in most of the descriptive studies. Figure 3a shows the most common network measures used to operationalize the theoretical mechanisms across the inferential articles, and figure 3b reflects the distribution of network measures used in descriptive articles. Degree centrality is the most common network measure in both inferential and descriptive articles. Among the inferential studies, the third most common measure is cohesion via diffusion. Interestingly, cohesion via diffusion is one of the least used measures in the descriptive articles. Moreover, in both types of articles, betweenness centrality is one of the most common measures. Across the inferential studies, the most common measure used for the operationalization of variables was "Other", we included in this category a large number of infrequently used



measures such as efficiency measure, participation coefficient, mutual dependence, etc. These measures were used two times or less across the inferential studies. We used the "Moderation effects" category whenever a hypothesis predicts that the impact of one variable in the network effectiveness is moderated by the magnitude of another variable [e.g., Shrestha (2018) states that the positive relationship between the number of collaboration partners and the possibility of being funded is stronger when the frequency of contact increases].

Figure 3a. Distribution of Network Measures Across Inferential Studies (56 articles)

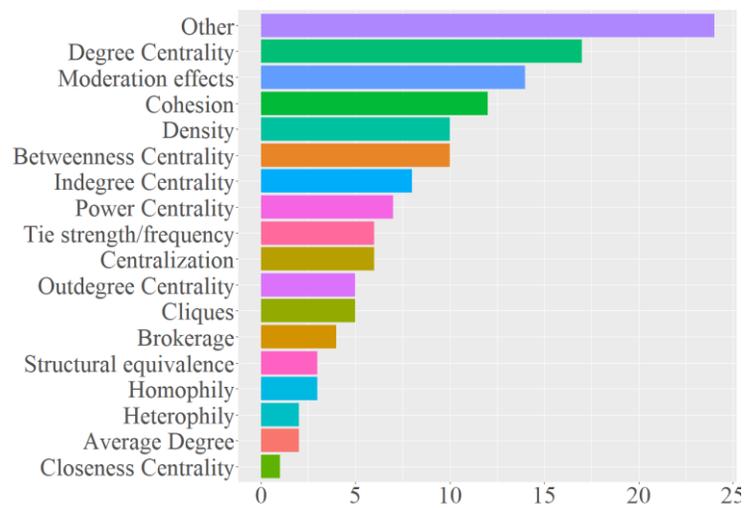

Figure 3b. Distribution of Network Measures Across Descriptive Studies (15 articles)

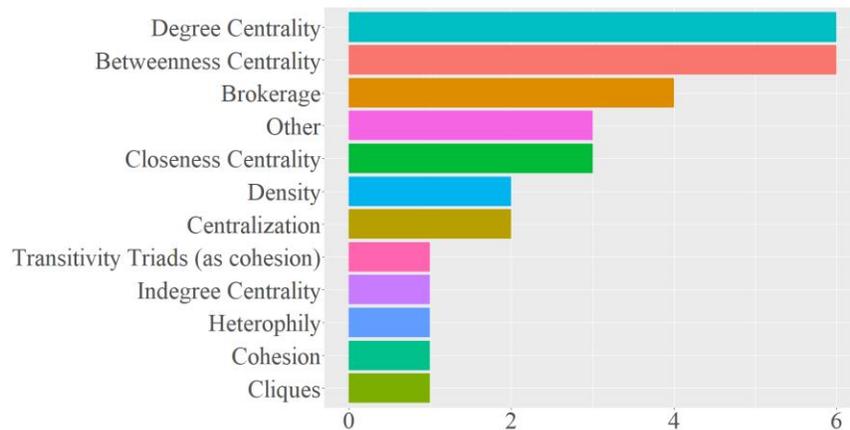



Below, we provide a summary of each of the social theories and structural arguments. To keep this paper relatively brief, we include only the social theories and structural arguments that were used in four or more hypotheses. For each, we describe the theory and its associated mechanisms along with the specific network measures used to capture the theory.

**Social Theories**

***Social Capital – Bridging***

  ***Theory and Mechanisms.*** We identified 18 hypotheses from 15 articles using social capital theory based on bridging relationships often discussed as structural holes and brokerage. Structural holes and positions of brokerage in networks are linked to network effects through two primary mechanisms: information and control (Borgatti & Foster, 2003; Burt, 1992; Lin, 1999). The general argument for the informational benefits of structural holes is largely based on the work of Granovetter (1983) and Ronald Burt (1992; 2005). He argues that actors that are directly connected tend to share similar opinions, ideas, and knowledge. Therefore, having ties to multiple individuals from the same group is inefficient as you can obtain roughly the same informational benefits from a single tie to the group (Burt, 1992). Access to diverse information and nonredundant ideas is best achieved by forming ties to individuals who are not connected. Individuals and organizations who span structural holes are viewed as entrepreneurs who obtain informational and vision advantages from their investment in bridging disconnected actors (Burt, 2005). Seven of the fifteen articles emphasize the role of informational benefits and the value of access to novel ideas and resources. For example, Jokisaari and Vuori (2010), in their research on job training programs in Finland, contend that brokerage provides individuals with early



access to information about new programs. Thus, these individuals are more likely to be early adopters of innovative practices.

In terms of control, structural holes and brokerage are thought to provide benefits by allowing the broker to manage the flow and movement of information. The benefits are captured by the concept of *tertius gaudens*, which translates to "the third who benefits" (Simmel, 1950). Monge and Contractor (2003) discuss two situations where benefits of control accrue to the broker (p. 144). First, brokers are uniquely positioned to profit when the two actors they broker are competing for the same resource. In this case, the broker can obtain a higher reward for that resource by playing the two bidders against each other. The second situation occurs when the two actors are seeking different resources, allowing the broker to operate as a mediator, controlling the information flow and ultimate resolution (Monge & Contractor, 2003, p. 144).

As noted by Borgatti and Foster (2003), if one can keep actors in the network disconnected, especially one's adversaries or competitors, then those actors are less able to coordinate against you (p.1003). Eight of the fifteen articles emphasize the role of control and influence over the movement of information and resources. For instance, Faulk et al. (2016) argue that organizational power results from an organization's ability to influence others by controlling the movement of information and the availability of opportunities in the network. Similarly, Marcum et al. (2012) argue that direct ties are insufficient to maintain authority in an interorganizational network. Rather, they suggest that indirect ties, where organizations can act as gatekeepers of information and resources, provide positions of authority over the actors they broker (p. 521).

*Measures.* Several structural and non-structural measures have been used to capture brokerage and structural holes in networks. The two most common measures are betweenness



centrality (Freeman, 1979) and brokerage scores (Gould & Fernandez, 1989). Other scholars

have used network constraint or efficiency as measures of structural holes. Each of these four

primary measures emphasizes an individual's bridging position between non-connected actors, a

structural role that aligns with both the mechanisms of information and control. However,

several scholars have utilized measures based on direct connections, such as degree centrality or

the average degree of actors in the network, to assess hypotheses about bridging or structural

holes. These measures have less alignment with the theoretical mechanisms posited by the

authors. Furthermore, not all scholars drawing on structural hole theory rely on structural

measures. Sandström and Carlsson (2008) argue that heterogeneity and diversity in networks

lead to improved outcomes. Thus, rather than focus on one's structural position to measure

information diversity, Sandstrom and Carlsson use the diversity of the actors (based on variation

in administrative affiliation) and interactions that cross boundaries within the network as

measures of structural holes.

### Social Capital-Bonding

***Theory and mechanisms.*** We identified eight hypotheses from seven articles using

bonding social capital theory. Unlike bridging social capital, bonding social capital refers to

connections and resources in densely connected homogeneous groups with similar backgrounds

or interests (Putnam, 2002). Bonding social capital is primarily described in two ways: the

strength of the ties connecting the actors and network cohesion. Tie strengths and network

cohesion affect networks by (a) directly encouraging cooperative behavior or (b) addressing

differences and mitigating the transaction costs in the coordination process. Strong ties provide

opportunities for network members to exchange information in a timely manner, build trust, and

establish social norms, therefore enhancing collaborative commitment and promoting



cooperative behavior (Coleman, 1988). Furthermore, a well-connected and closed network structure makes it easy to verify the credibility of information, leading to a reduction in uncertainty and prevention of defective behavior (Arnold et al., 2017; Berardo & Scholz, 2010).

*Measures.* Among the articles reviewed, researchers operationalized the strengths of ties as the frequency of interactions between two nodes. For instance, Lee (2013) used communication frequency to measure tie strength and examined its impacts on perceived e-government effectiveness. Stronger ties between program units and IT units, between service programs and outside IT vendors contributed to e-government effectiveness. Similarly, Lee and Kim (2011) used communication frequency to measure employee's tie strength and studied their relationship with affective commitment. Network cohesion has been operationalized as the structural measures of transitive triads, clustering coefficients, and subgroup cohesion (e.g., Arnold et al., 2017; Shrestha, 2013; Yi, 2018). A high proportion of transitive triads have been found useful in building shared identity, offering support, and making it difficult for outside interventions to influence network members (Arnold et al., 2017). A high clustering coefficient (the proportion of links among an ego's alters) suggests a high level of redundancy of linkages exists in the network, which can help address differences or conflicts among members (Yi, 2018). Shrestha (2013) proposed a similar measure, "subgroup cohesion"—the percentage of the organization's alters or partners working on the same projects—to study its impacts on securing program funds for the collaborative Rural Water Supply and Sanitation Program (RWSSP) in Nepal.

### Social Capital – Trust

*Theory and Mechanisms.* Six hypotheses across four papers focused on the role of trust in shaping network effects. A basic definition of trust is the expectation that actors will take each



other's interests into account when making decisions or taking action (Lin, 2001, p. 147). Trust is seen as a critical element to the basic functioning of society and the multitude of formal and informal relationships that exist (Simmel, 1950). Trust is tightly linked with the concept of social capital. Because social capital emerges through one's relations and the resources made available from those relations, trust is viewed as a component or attribute of social capital (Kadushin, 2012). As such, "Networks are both indicators of social capital and also are a process that leads to social capital" (Kadushin, 2012, p. 177). In the articles reviewed, the trust component of social capital is used as a predictor of network effects. The implications of trust examined in the articles range from charitable giving (Markovic, 2017) to obtaining project funds for community development (Shrestha, 2013).

*Measures.* Trust is an inherently difficult concept to measure. Authors most often use survey questions to obtain perception-based measures of trust. For example, Hawkins (2010, p. 260) developed an additive index based on respondents' level of agreement with a number of items, such as: "local government officials from different jurisdictions trust one another." Herzog and Yang (2018) and Markovic (2017) used a similar approach but relied on a single survey item. Shrestha (2013) examines the role of social trust in a community on that community's success in obtaining financial resources. He relies on a measure of civic engagement to capture social trust. Based on Putnam, Leonardi, and Nanetti (1993), civic engagement was measured by the proportion of households in the community who were members of local organizations.

**Policy Diffusion/Social Influence Theories**

*Theory and Mechanisms.* Ten hypotheses across seven articles were supported by Policy Diffusion/Social Influence theories. There are two primary mechanisms associated with the diffusion of attitudes and innovation (Burt, 1987). One is cohesion or peer influence. The idea is



that through repeated communication with network members, an ego is more likely to gain information regarding the new behavior, policy or attitude from alters that have already adopted (Jokisaari & Vuori, 2010). Furthermore, the larger the number of actors adopting a behavior or attitude in one's immediate network, the more legitimate the behavior or attitude is perceived socially (DiMaggio & Powell, 1983). The other mechanism is structural equivalence, which means that two nodes have the same connection patterns with other nodes in a network (regardless of whether the two nodes are directly connected). Having the same connections means that they pick up the same information from other nodes in the network, including new behaviors or innovations. Moreover, the competition between two structurally equivalent nodes is likely to motivate the adoption of innovations (Burt, 1987; Valente, 2010). If one node adopts an innovation, the other is very likely to adopt it in order to stay competitive or legitimate in the eyes of other nodes in the network.

   *Measures.* We observe multiple ways that scholars operationalize diffusion and influence. A popular way to operationalize social influence is to examine the number of direct contacts that have adopted a certain behavior or attitude (Jokisaari & Vuori, 2010; Kammerer & Namhata, 2018). Interactions with direct contacts expose an actor to this new behavior or attitude; the more contacts adopt this behavior, the more socially appealing it becomes. Building on this approach, Siciliano and Thompson (2018) constructed a weight matrix that adjusts the relative influence of any given alter's organizational commitment based on the strength of the ego's connection to that alter relative to the strength of the ego's other connections. An alternative way, based on the structural equivalence mechanism, is to look at whether two actors have the same patterns of connections with other actors (Cao & Prakash, 2011). In the context of international trade, competing actors will tend to emulate each other if they have the same



patterns of connections with others because they want to be perceived to be at least equally legitimate. In this approach, social influence is caused mainly by perceived legitimacy.

The above two approaches to operationalize social influence are structure-based. The transfer of agents across networks also provides opportunity for the diffusion of policies or innovations. Yi, Berry, and Chen (2018) examined how the transfer of agents from one organization to another may bring with them the practices and ideas from their first position. As a result, the performance of these organizations may be correlated.

### Resource dependence theory

*Theory and Mechanisms.* Resource dependence theory is used to support six hypotheses across five articles. Resource dependence theory characterizes organizations as open systems that depend on their environments for critical resources (Pfeffer & Salancik, 1978). Therefore, maintaining stable supply of critical resources largely determines how organizations interact with other organizations.  For example, organizations may take strategies such as a merger or diversification of supplies to reduce their dependence on certain organizations. Researchers can thus study organizations' decision to build or dissolve relationships from a resource dependence perspective. The dependence of an ego on an alter for resources is the basis of power that the alter has over the ego (Emerson, 1962). Relationships within networks are not always created equal. Some organizations rely more heavily on others for critical resources, creating imbalances in power relationship between organizations (Emerson, 1962). For instance, if a nonprofit organization relies on a single agency for funding, then the nonprofit is vulnerable to the funder's changing funding demands (AbouAssi & Tschirhart, 2018).

*Measures.* Resource dependence theory is thus often used to study how organizations manage relationships within networks, but there are various ways to operationalize key



constructs: power or dependence. Provan, Huang, and Milward (2009) took a structural approach and used organizational centrality in a network to measure the influence or power of an organization. The idea is that "central organizations can maintain a 'gatekeeping' role in the network, controlling access to valued resources" (p.877). Varone, Ingold, and Jourdain (2017) used the number of groups in a lobbying coalition as a measure of resources. The idea is that the more groups, the more resources can be mobilized. For organizations that rely on others for resources, measuring the degree of dependence is an important question. AbouAssi and Tschirhart (2018) used the percentage of a nonprofit's funding that comes from a funding agency to measure the degree of dependence.

Some scholars used a survey approach to measure resources. For example, Lee and Lee (2018) study how social interactions derived from resource exchanges are related to trust among members. They measure resource exchange relationships between organizations by asking respondents whether or not their organizations shared six types of social interactions derived by resource-dependence relationships with all other organizations. Resh, Siddiki, and McConnell (2014) hypothesized that interorganizational relationships based on resource exchanges facilitate organizational learning. Their measurement is based on surveys that ask respondents to evaluate the importance of financial resources, and the influence that an alter has in determining ego's coordination with the alter.

**Structural Arguments**

*Centrality*

  ***Theory and mechanisms.*** We identified 45 hypotheses across 24 articles that relied on the centrality of actors for the prediction of network effects. Centrality is one of the most common nodal measures used in network analysis for understanding the relevance of a node



based on its position in a network (Wasserman & Faust, 1994). As noted in the social theories above, centrality has been used to test several mechanisms that increase network effectiveness. Here, rather than centrality being the chosen measure to capture a particular mechanism, authors rely on structure-based arguments to predict network effects and consider centrality as the driver of better network outcomes. At the network level, centrality has also been tested to identify if network effectiveness increases when the network is organized around one or few actors.

*Measures.* We found nine hypotheses that focus on how the central position of actors within the network facilitates access to novel and broader information that leads to network effectiveness. This mechanism is similar to the one associated with social capital bridging theory. However, we coded these hypotheses under the centrality argument category because authors rely mainly on the benefits of centrality as a structure for network outcomes and not on arguments based on social capital bridging theory. Scholars operationalize this argument through different centrality measures like degree centrality, closeness centrality, and betweenness centrality. Some scholars use degree centrality, predicting that having more ties offers the actor greater access to broader and complex information and resources, leading to increases in network effectiveness (Arnold et al., 2017; Lee, 2013; Pappas & Wooldridge, 2007; Schalk et al., 2009; Scott & Thomas, 2017; Shrestha, 2013; Siciliano, 2017). Closeness centrality was used to test that in addition to having more ties, the shorter distance between nodes, the faster the information flows (Borgatti et al., 2018), leading to greater effectiveness.

Thirteen hypotheses across six articles test how direct control and influence impact network effectiveness. Researchers use different centrality measures to operationalize control over other actors like degree centrality (Dekker et al., 2010; Marcum et al., 2012; Wong & Boh, 2014), indegree, and outdegree centrality (Hu & Kapucu, 2016). The main argument posited by



these authors is that central actors in the network have better capability to use information for coordination purposes and develop managerial innovative skills. In two articles, scholars are interested in testing the impact of a node's position in the network for access to better resources. They use betweenness centrality to test how optimally positioned actors control the flow of information and resources in the network (Koliba et al., 2017) or access to foundation grants (Faulk et al., 2016).

Four hypotheses focus on testing indirect control over other actors. Scholars operationalized this mechanism using eigenvector centrality and power centrality. Eigenvector centrality assesses the indirect influence over other actors when nodes are connected to nodes that are also well connected (Bonacich, 1987; Borgatti et al., 2018; Wasserman & Galaskiewicz, 1994). For instance, researchers use this measure to test if managers' skills in communicating new strategic actions are related to the extent to which managers are connected to well-connected actors (Marcum et al., 2012; Pappas & Wooldridge, 2007). Power centrality (also referred to as beta centrality) is similar to eigenvector centrality in that a node's status depends on the status of the other nodes to whom it is connected (Bonacich, 1987). Arnold et al. (2017) test if groups promoting legislation are more successful when the actors are connected to other well-connected municipal actors.

Four hypotheses use centrality measures of degree and indegree as the direct driver of different organizational behaviors. Lee and Kim (2011) posit that more relationships (i.e., higher degree) may increase a sense of belonging to the organization and tested the extent to which having more ties affects employees' level of affective commitment. Vardaman et al. (2012) predict that a higher number of received friendship and advice ties is associated with higher abilities to adapt to organizational change.



Finally, centrality is also operationalized at the network level to test the impact on network effectiveness when the network is organized around a particular actor or actors (Provan, K. & Milward, 1995). Based on this argument, six hypotheses used centralization measures to predict that integration and coordination lead to greater network effectiveness (Hawkins, 2010; Markovic, 2017; Raab et al., 2015), even in the presence of conflicting goals (Akkerman et al., 2012). Klaster et al. (2017) state that centralization in the short run leads to greater effectiveness because the implementation process starts sooner due to a shorter time of decision making and discussion processes.

### *Structural embeddedness*

*Theory and mechanisms.* Eight hypotheses from five articles were built upon structural embeddedness. Different from relational embeddedness that focuses on the quality or strengths of relations, structural embeddedness focuses on the pattern of relations in the network (Granovetter, 1992). Some scholars embraced a definition of structural embeddedness as the extent to which "a dyad's mutual contacts are connected to one another" (Granovetter, 1992, p. 35). Others used structural embeddedness to refer to the extent to which a node takes a central position in the network (e.g., Provan et al., 2009) or whether a node belongs to a cohesive subgroup (e.g., Schalk, Torenvlied, & Allen, 2010).

*Measures.* In the five articles reviewed, different measures of structural embeddedness have been proposed, including degree centrality, power centrality, density and centralization, cliques, and the number of shared collaborators between organizations (Provan et al., 2009; Sandström & Carlsson, 2008; Schalk et al., 2009; Villadsen, 2011). For instance, in a study of the collaboration among rural economic development organizations, Ofem and his colleagues (2018) argued that structural embeddedness--"the extent to which two organizations share



multiple collaborators" -- influences their collaborative outcomes (p. 1116). Building structural embeddedness through common collaborators not only allows the two member organizations to verify easily important information and reduce risks of coordination, and it also enhances network partners' commitment and cooperative behavior (Ofem et al., 2018).

Network centrality also exposes the node to "the normative framework of what constitutes proper or legitimate behavior," thereby placing institutional constraints on its behavior in the network (Villadsen, 2011, p. 579). Villadsen (2011) suggested that the structural embeddedness of Danish mayors, measured by their degree centrality in the policy network, influences policy isomorphism in municipalities. Provan, Huang, and Milward also operationalized structural embeddedness as centrality in their research of mental health service networks (2009). But they argued that Bonacich power centrality is a better indicator of structural embeddedness than degree centrality because power centrality considers both the direct ties a focal node has and the position of alters in a network.

Structural embeddedness has also been operationalized as "cohesive subgroup membership" (Schalk et al., 2009, p. 630). Schalk et al. (2009) argued that strong and close relations within cohesive subgroups are important for trust building and reducing transaction costs. In their study of Dutch higher education, they found that colleges' affiliation with cohesive subgroups (cliques), rather than their degree centrality, contributes to their performance which was measured by students' evaluations.

### Density

***Theory and mechanisms.*** We found five hypotheses across four articles that use density arguments to predict network effectiveness. Density is a network-level variable that is measured as the ratio of the number of ties in the network to the number of possible ties (Borgatti et al.



2018). Dense networks are often viewed as an indicator of a "bonding social capital", trust, and cooperation (Gauthier, 2020, p. 468; Lin 1999; Coleman, 1994). Scholars offer several different mechanisms by which densely connected networks lead to network effectiveness. Hawkins (2010) uses density to test if greater policy network cohesion increases collaboration with other local governments for economic development purposes through the possibility of a joint venture formation. Dekker et al. (2010) test if behaviors like trust, support, and cooperation are more likely to be found in dense networks. Lee (2013) argues that actors embedded in dense networks experience a higher perception of e-government effectiveness for public service delivery.

*Measures.* As noted above, density is the proportion of ties present out of the total possible ties (Borgatti et al., 2018). While density is a straightforward measure, there are some important caveats when comparing the level of interconnectedness among networks of different sizes. As network size changes, similar levels of density do not necessarily reflect the same levels of connectedness among the network members (Bodin et al., 2017; Borgatti et al., 2018; Sandström & Carlsson, 2008). For instance, if we compare a network with 10 members against a network with 100, where each member has two ties in both networks, the density of the smaller network is 0.22, whereas the density of the larger network is 0.02. In this regard, as Borgatti et al. (2018) state, when comparing networks of different sizes, we need to consider that levels of density tend to be lower in large networks because as the network grows, it is difficult for the nodes to create ties with of the same percentage of the members.

### Other & Multiple Categories

Eight hypotheses across four articles were coded as "Other" in the theory category as the theories were not frequently used in the network literature. Some examples are entropy theory, cooptation theory, production theory, systems theory, and the advocacy coalition framework. We also found



five hypotheses across four articles that used multiple theories or structural mechanisms for support. For example, Akkerman et al. (2012) relied on social capital and resource dependence theory to test if participation in collaborative subnetworks (cliques) affects performance due to resource exchange and the building of interorganizational trust. Similarly, Provan et al. (2013) use arguments from policy diffusion and homophily theories to predict that frequent and more intensive ties between similar organizations will allow for greater information exchange and increase the awareness of new practices.

*NA category*

We coded 89 hypotheses across 31 articles as "NA" in the theory category when authors framed their hypotheses based on the specific research context instead of a social theory or structural argument. As Siciliano et al. (2021) state, many context-framed hypotheses in public administration and public policy are related to the fact that these fields are not only scientific but also applied (p. 13). For instance, Resh et al. (2014) test if learning is higher in collaboration venues in which government actors are more central or in venues with higher levels of trust. Instead of framing the hypotheses according to a specific social theory, the authors relied on general findings of the collaborative governance, public management, and public policy literatures. Similarly, Faulk et al. (2016) test the extent to which nonprofit organizations gain more grants if they have board interlocks with foundations. The authors frame this hypothesis based on the nonprofit management literature and on empirical findings related with factors that increases the probability of nonprofit organizations accessing  foundation grants.

## Discussion and Conclusions

This systematic review aimed to understand the mechanisms and measures used by public administration and public policy scholars studying network effects. Based on our systematic



literature review of 40 journals in public administration and policy, we present five main topics and challenges to our field's scholarship on network effects: (i) limited use of social theory for explanation, (ii) structure as mechanism versus measure, (iii) inconsistent use of network measures associated with particular theories, (iv) multiple mechanisms associated with any single theory, and (v) ability to *capture network dynamics and coevolution.*

**Limited Use of Social Theories**

One of the main findings from our review is the limited use of social theories to justify hypotheses that predict network effects. As shown in figure 4, out of the 223 hypotheses we extracted from the 56 inferential articles, only 71 (32%) were justified based on specific social theories. The most common social theories used by network scholars were social capital-bridging (18 hypotheses) and policy diffusion (10 hypotheses). Forty percent of hypotheses relied on either the research context for hypothesis development and/or failed to provide clear theoretical justification. These studies did not address the mechanisms through which network effects were produced. In comparison with the use of social theories, we found that 62 of the inferential hypotheses (28%) were framed based on structural arguments and the most common structural argument was centrality (45 hypotheses). This leads to our second challenge, almost the same number of network effects hypotheses rely on structure-based arguments rather than theory for the explanation of specific mechanisms at work that lead to better outcomes.



Figure 4. Use of Theory Across Effectiveness Literature

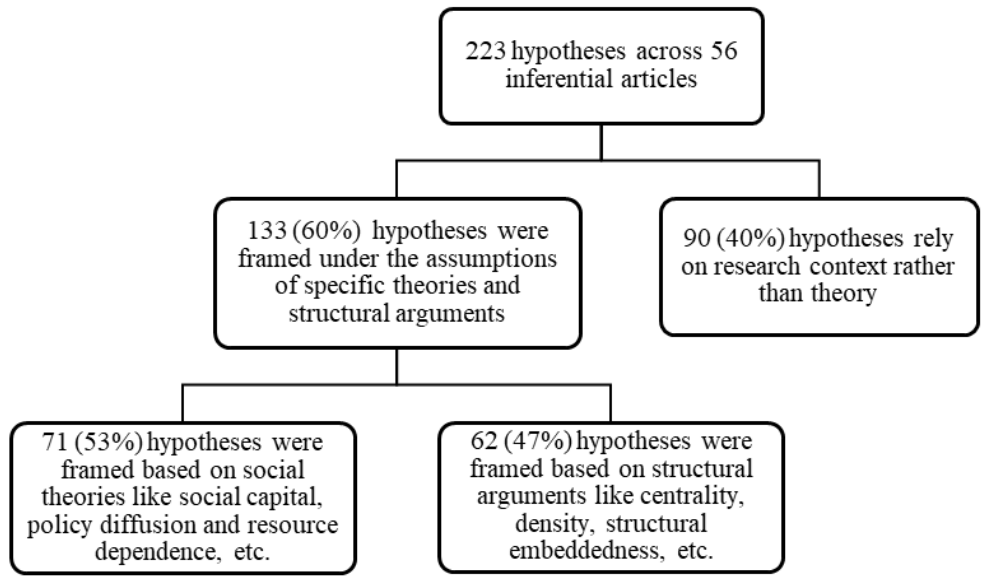

**Structure as mechanism versus measure**

Given the use of both theoretical justifications and structure-based arguments for network effects, an important question arises as to whether network consequences derive from specific structural arrangements or if structure only serves as a proxy measure of some mechanisms. Take the research on social capital and brokerage as an example. Networks effects have been posited to arise due to both the structural arrangement and the attributes of the individuals (e.g., Ter Wal et al., 2016). We see this tension between mechanisms and measures displayed in the selection of the network variables used to capture social capital via brokerage. Some authors rely on purely structural measures (e.g., constraint, efficiency), while others rely on non-structural measures such as the diversity of the actors. However, one could have strong, close ties to a set of diverse actors or bridging ties to actors who may be very similar (say all from the same administrative unit).



Burt (2004) states that "Networks do not act, they are a context for action" (p. 354). In this sense, if we rely on structure-based arguments to predict network effects, we assume that the structure itself leads to better outcomes, leaving behind any further explanation of the actual mechanisms at work. Further discussion of network structure as mechanism vs a measure is needed to create shared knowledge about the exogenous and endogenous drivers of greater network effectiveness. This distinction between mechanism and measure is crucial for instances where researchers and practitioners seek to intervene in networks in ways that may potentially alter their shape.

**Inconsistent use of network measures**

Another challenge we found in the effectiveness literature is the operationalization of mechanisms through network measures. For instance, degree centrality was the most common measure used across inferential and descriptive articles. However, this measure was operationalized to test different mechanisms associated with centrality like access to broader information, coordination of activities, direct control and influence, affective commitment, and interpretation of organizational change. The challenge we find with the use of measures like degree centrality is that the same measure is used to test different mechanisms, potentially hindering our ability to identify the specific mechanisms at work.

We also found a lack of agreement on how specific mechanisms should be operationalized. For instance, one approach to measure diffusion is through the number of direct contacts that have adopted  certain behaviors or attitudes. However, other scholars suggest that this approach does not consider the possibility that different strengths of ties may have differential influences on adoption. Similar challenges also existed for operationalizing mechanisms associated with embeddedness and bridging social capital.



Table 2 identifies the network measures that are used to test the associated mechanisms across the social theories and structural arguments discussed in the previous section. The rows refer to the social theories and structural arguments we described in the findings section, and the columns include the network measures used to operationalize the mechanisms associated with each theory and argument. A check means that a certain structure measure is used by a theory, and a letter "X" means that the use of that network measure is not justified according to the literature, and that there are other measures more suitable to test the mechanisms associated to that social theory or structural argument. For instance, according to social capital literature the mechanisms associated with bridging are better operationalized through network measures like betweenness centrality and brokerage instead of degree centrality or average degree. Additionally, betweenness centrality could be a better measure for explaining the impact of control over others on network performance than degree centrality (Wasserman & Faust, 1994). The social capital-trust theory has no network measure assigned to it. Our findings reported that scholars use perception-based measures such as the level of agreement or civic engagement to operationalize trust.

Table 2. Network Measures and Theories

| Theory \ Network Measure | Degree Centrality | Outdegree Centrality | Indegree Centrality | Power Centrality | Betweenness Centrality | Eigenvector Centrality | Closeness Centrality | Average Degree | Centralization | Density | Transitivity | Heterophily | Cliques | Cohesion (peer influence) | Tie Strength | Structural Equivalence | Brokerage |
|---|---|---|---|---|---|---|---|---|---|---|---|---|---|---|---|---|---|
| Policy Diffusion/Social Influence | | | | | | | | | | | | | | ✓ | | ✓ | |
| Resource Dependence | ✓ | | | ✓ | | | | | | | | | | | | | |
| Social Capital-Bridging | X | | | | ✓ | | | X | | | ✓ | | | | | | ✓ |
| Social Capital-Bonding | | | | | | | | | | ✓ | | | | | ✓ | | |



| | | | | | | | | | | | | | | | | |
|---|---|---|---|---|---|---|---|---|---|---|---|---|---|---|---|---|
| Social Capital-Trust | | | | | | | | | | | | | | | | |
| Structural Argument-Centrality | ✓ | ✓ | ✓ | ✓ | ✓ | ✓ | ✓ | | ✓ | | | | | | | |
| Structural Argument-Density | | | | | | | | | | ✓ | | | | | | |
| Structural Argument-Structural Embeddedness | X | | ✓ | | | | | | | | | | ✓ | | | |

## Multiple mechanisms associated with a single theory

A fourth challenge is that even for a single theoretical argument, several mechanisms may be operative. Social theories are inherently complex and are often comprised of multiple mechanisms. For example, with regard to social capital bridging, the advantages to brokers are argued to result from information diversity, information control, or both. While some authors rely on information diversity as the driver of performance and others on information control, the structural measures used to capture these processes are often identical. Developing designs to isolate the operative mechanisms is needed to better understand how social capital from structural holes and brokerage positions translate into better outcomes. Similarly, research relying on social capital-bonding as a performance driver focuses mainly on the structural benefits of being in a cohesive network, but alters' attributes, support, and resources also have important implications on performance (Lin, 1999).

This challenge is particularly acute when scholars rely on structural arguments to support their hypotheses. For instance, in our analysis of the hypotheses relying on the structural argument of centrality, we found that authors associate centrality with greater access to information, acquisition of resources, speed of information flow, ability to establish control and influence in the network, as well as affective feelings of belonging and commitment. Density is another example of a structural argument used to test different mechanisms like trust, support, and collaboration. One example where multiple mechanisms have been tested and disentangled is concerning policy diffusion. For instance, scholars tested the mechanisms of peer influence



and structural equivalence and posited two different processes that may lead members of a network to have similar behaviors or adoption times and tested those processes against each other (Burt, 1987). Such theory-informed measures can guide authors interested in testing and comparing the impact of different mechanisms against each other.

**Network dynamics and coevolution**

Networks are not static. The ties actors form may dissolve and at times reappear. As the relationships change, so too may the observed behavior of the actors in the network. For instance, one of the challenges with identifying the implications of trust in networks is that while networks can enable the establishment of trust, trust also facilitates the development of network relations. The potential tautology among trust and networks was a primary source of criticism of Putnam's book Bowling Alone (2000). Critiques have highlighted the issue of logical circularity. Portes (1998, p. 19) states "As a property of communities and nations rather than individuals, social capital is simultaneously a cause and an effect. It leads to positive outcomes, such as economic development and less crime, and its existence is inferred from the same outcomes." A similar challenge is faced by scholars who analyze the role of trust in producing network effects. Those network effects, such as higher funding levels or willingness to donate to charities, are driven by trust but also those successes or actions further the development of trust. Trust develops over time as actors exchange resources and maintain promises; trust is continually confirmed, additional interactions and exchanges are more easily accomplished (Koppenjan & Klijn, 2004). Trust, as with many other feelings and behaviors, coevolves alongside the network.

This challenge raises important questions for research on network effects. With cross-sectional data, it is impossible to differentiate between selection and influence (Shalizi and Thomas 2010). Thus, when individuals with similar attitudes or behaviors are connected in a



network, it is unknown if the similarity in behavior drove their social relationship or if the social relationship produced their similar behavior. When attempting to distinguish between selection and influence effects in networks, longitudinal data can help. However, only two out of seven studies of policy diffusion and social influence use longitudinal data.

By analyzing 74 articles related to network effects we identify the most common social theories and structural arguments used to justify hypotheses and mechanisms associated with effectiveness. Overall, more theory-based research is needed as a way of advancing knowledge within the field of how network structure and actor attributes influence effectiveness. Similarly, considering the high number of structure-based hypotheses in the field, there needs to be further discussion regarding the convenience of using structural arrangements as mechanisms instead of as network measures to explain network outcomes. In general, the findings of this article could assist scholars in identifying different theories that can be used to justify various mechanisms associated with network effectiveness. In addition, this analysis could assist network researchers in building agreement on how to operationalize mechanisms related to network effectiveness framed under a specific theory. Finally, further examination is required to understand the extent to which the hypotheses used in the network effectiveness literature are supported, and if so, under what specific circumstances the mechanisms at work have a positive or negative impact on effectiveness.

**Notes**

[1] Besides "network effects", in this manuscript we also use the terms "network effectiveness" and "network performance" as they are used in the reviewed articles. "Network effects" refers to network outcomes, "network effectiveness" is used when discussing the network effectiveness



literature or network research in general, and "network performance" refers to metrics or individual/group level outcomes.

[2] We followed the approach of Kapucu et al. (2017) and included in our search 39 public administration and policy journals previously identified by scholars as relevant journals in the field (Bernick and Krueger 2010; Forrester and Watson 1994). We added *Public Management Review*, given the number of network studies published by this journal. A list of the 40 journals and the number of articles extracted from each journal is available in an online appendix.